\documentclass[aps,prb,groupedaddress,showpacs,twocolumn]{revtex4}
\usepackage{graphicx}
\usepackage{amsmath}
\usepackage{bbm}
\usepackage{xspace}
\usepackage{color}

\newcommand{\psystem}{physical system}

\newcommand{\se}{Sec.\@\xspace}

\newcommand{\ie}{i.\thinspace{}e.\@\xspace}

\newcommand{\ptl}{\partial}

\newcommand{\PDF}[2]{\frac{\ptl\, #1}{\ptl\, #2}}

\newcommand{\ve}[1]{{\bf #1}}
\newcommand{\mat}[1]{\mathsf{#1}}

\newcommand{\eq}[1]{Eq.\thinspace{}(\ref{#1})}
\newcommand{\eqq}[2]{Eqs.\thinspace{}(\ref{#1}) and (\ref{#2})}

\newcommand{\tab}[1]{Tab.\thinspace{}\ref{#1}}

\newcommand{\fig}[1]{Fig.\thinspace{}\ref{#1}}
\newcommand{\figg}[2]{Figs.\thinspace{}\ref{#1} and \ref{#2}}
\newcommand{\fc}[1]{({#1})}
\newcommand{\figc}[2]{Fig.\thinspace{}\ref{#1}\thinspace{}\fc{#2}}
\newcommand{\figcc}[3]{Figs.\thinspace{}\ref{#1}\thinspace{}\fc{#2} and \fc{#3}}

\newcommand{\Tr}{\mbox{Tr}}

\def\ket#1{\mathinner{|{#1}\rangle}}

\begin{document}


\title{Benchmarking the variational cluster approach by means of the one-dimensional Bose-Hubbard model}


\author{Michael Knap}
\email[]{michael.knap@tugraz.at}
\affiliation{Institute of Theoretical and Computational Physics, Graz University of Technology, 8010 Graz, Austria}
\author{Enrico Arrigoni}
\affiliation{Institute of Theoretical and Computational Physics, Graz University of Technology, 8010 Graz, Austria}
\author{Wolfgang von der Linden}
\affiliation{Institute of Theoretical and Computational Physics, Graz University of Technology, 8010 Graz, Austria}


\date{\today}

\begin{abstract}
Convergence properties of the variational cluster approach with
respect to the variational parameter space, cluster size, and boundary
conditions of the reference system are investigated and discussed for
bosonic many-body systems. 
Specifically, 
 the variational cluster approach is applied to the one-dimensional
 Bose-Hubbard model, which exhibits a quantum phase transition from
 Mott
to superfluid phase. In order to benchmark the variational cluster
approach, results for the phase boundary delimiting the first Mott
lobe are compared with essentially exact density matrix
renormalization group data. Furthermore, static quantities, such as the
ground state energy and the one-particle density matrix are
compared with high-order strong coupling perturbation theory
results. For reference systems with open boundary conditions the
variational parameter space is extended by an additional variational
parameter which allows for a 
more uniform particle density on the reference
system and thus drastically improves the results. It turns out that
the variational cluster approach yields accurate results with
relatively low computational effort for both 
the phase boundary as
well as the static properties of the one-dimensional Bose-Hubbard model,
even at the tip of the first Mott lobe where correlation effects are
most pronounced. 
\end{abstract}


\maketitle

\section{\label{sec:introduction}Introduction}
Seminal experiments on ultracold gases of atoms trapped in optical
lattices\cite{jaksch_cold_1998, greiner_quantum_2002,
  bloch_many-body_2008} have lately turned the spotlight of scientific
research interest on the Bose-Hubbard (BH)
model\cite{fisher_boson_1989} and its variants. The BH model exhibits
a quantum phase transition from localized Mott phase to delocalized
superfluid phase. The Mott phase is characterized by integer particle
density, a gap in the single-particle spectral function and zero
compressibility.\cite{fisher_boson_1989} The regions in the phase
diagram where the ground state of the BH model is in a Mott state are
termed Mott lobes. The evaluation of the boundaries delimiting the
Mott phase and other physical quantities is very demanding for the
one-dimensional BH model as correlation effects are most important in
the low-dimensional case. This is reflected in the special shape of
the Mott lobes. Particularly, the lobes are point shaped and a reentrance behavior can be
observed\cite{khner_one-dimensional_2000, khner_phases_1998, Elstner_Dynamics_1999, koller_variational_2006} in contrast to the mean
field results.\cite{fisher_boson_1989, sheshadri_superfluid_1993,
  van_oosten_quantum_2001, sachdev_quantum_2001,
  menotti_spectral_2008} 

In the present paper, we benchmark the variational cluster
approach\cite{potthoff_variational_2003} (VCA) using the
one-dimensional BH model. VCA is based on the self-energy functional
approach\cite{potthoff_self-energy-functional_2003-1,
  potthoff_self-energy-functional_2003} and is a variational
extension of the cluster perturbation
theory\cite{snchal_spectral_2000, snchal_cluster_2002} (CPT), where
the \psystem{} is decomposed into clusters and the inter-cluster
hopping is treated perturbatively. It is crucial to investigate the
convergence properties of VCA in dependence of the variational
parameter space, the size of the clusters and the boundary conditions
used for the cluster Hamiltonian. The motivation for this research has
been the increasing interest in strongly correlated bosonic systems
such as ultracold gases of atoms trapped in optical
lattices\cite{jaksch_cold_1998, greiner_quantum_2002,
  bloch_many-body_2008} and light-matter
systems.\cite{greentree_quantum_2006, hartmann_strongly_2006,
  hartmann_quantum_2008} Lately, cluster methods have been benchmarked
for fermionic systems in Ref.~\onlinecite{balzer_mott_2008}, where the
authors used the one-dimensional fermionic
Hubbard-Model,\cite{hubbard_electron_1963} which can be solved exactly
by means of the Bethe ansatz,\cite{lieb_absence_1968} in order to test
the achievements of VCA. However, it remains an open question how VCA
performs in the case of bosonic systems. Unfortunately, an exact
solution of the one-dimensional BH model does not exist, as each
lattice site can be occupied by infinitely many bosonic particles. Yet
essentially exact density matrix renormalization group (DMRG) results
for the phase boundary delimiting the first Mott lobe are
available,\cite{khner_one-dimensional_2000} and static properties, such
as the ground state energy and the one-particle density matrix,
have been evaluated using strong-coupling perturbation theory of high
order.\cite{damski_mott-insulator_2006,teichmann_process-chain_2009}
In the present paper we discuss the convergence properties of VCA for
these physical quantities. 

The outline of this paper is as follows. In \se~\ref{sec:model}, the BH
model is introduced. Section~\ref{sec:method} describes the most
important aspects of VCA. In this section, the ordinary variational
parameter space of the BH model used for cluster Hamiltonians with
open boundary conditions is extended by an additional variational
parameter, which allows for a better distributed particle density
within the cluster and therefore drastically improves the results. The
convergence properties of VCA for the phase boundary of the first Mott
lobe are investigated in \se~\ref{sec:spectral} by comparing the VCA
results for distinct sets of variational parameters and cluster sizes
with DMRG data from
Ref.~\onlinecite{khner_one-dimensional_2000}. Additionally,
single-particle spectral functions and densities of states are
evaluated and their dependence on the variational parameter space is
discussed. In \se~\ref{sec:static} the ground state energy and the
one-particle density matrix are calculated and compared with
strong coupling results of high order from
Ref.~\onlinecite{damski_mott-insulator_2006}. Finally,
\se~\ref{sec:conclusion} concludes and summarizes our findings. 

\section{\label{sec:model}The Bose-Hubbard model}
The BH Hamiltonian\cite{fisher_boson_1989} is given by
\begin{equation}
 \hat{H}=-t \sum_{\left\langle i,\,j \right\rangle} b_i^\dagger \, b_j
+ \frac{U}{2} \sum_i \hat{n}_i\left(\hat{n}_i-1 \right) - \mu\,\hat{N}_p \; \mbox{,}
 \label{eq:bhm}
\end{equation}
where $b_i^\dagger$ and $b_i$ are bosonic creation and
annihilation operators at lattice site $i$, $t$ is the hopping
strength between two adjacent sites, $U$ is the on-site repulsion and
$\mu$ is the chemical potential, which controls the total particle
number $\hat{N}_p=\sum_i \hat{n}_i = \sum_i b_i^\dagger \, b_i$. The
first term in \eq{eq:bhm} is considered as a sum over nearest
neighbors $j$ of site $i$. In the calculations and the forthcoming
discussions we use the on-site repulsion $U$ as unit of energy.

\section{\label{sec:method}The variational cluster approach}

The variational cluster approach has been originally proposed for
fermionic systems in Ref.~\onlinecite{potthoff_variational_2003} and
has been extended to bosonic systems in
Ref.~\onlinecite{koller_variational_2006}. The main idea of VCA is
that the grand potential $\Omega$ is expressed as a functional of the
self-energy $\mat{\Sigma}$. At the stationary point of the self-energy
functional $\Omega[\mat{\Sigma}]$ Dyson's equation for the Green's
function is recovered. Unfortunately the functional
$\Omega[\mat{\Sigma}]$ cannot be evaluated directly, as it contains
the Legendre transform of the Luttinger-Ward
functional.\cite{lu.wa.60,potthoff_self-energy-functional_2003-1}
However, the latter just depends on the interaction part of the
Hamiltonian and thus it can be eliminated by comparing
$\Omega[\mat{\Sigma}]$ with the functional
$\Omega^\prime[\mat{\Sigma}]$ of a simpler, exactly solvable system
$\hat{H}^\prime$, which shares the interaction part with the 
\psystem{} $\hat{H}$. The system $\hat{H}^\prime$ is referred to as
reference system. With these considerations one obtains for bosonic
systems\cite{koller_variational_2006} 
\begin{alignat}{2}
 \Omega[\mat{\Sigma}] &= \Omega^\prime[\mat{\Sigma}] &&- \Tr\,\ln(-(\mat{G}_0^{\prime\,-1}-\mat{\Sigma})) \nonumber \\
  & &&+ \Tr\,\ln(-(\mat{G}_0^{-1}-\mat{\Sigma})) \; \mbox{,}
 \label{eq:om}
\end{alignat}
where quantities with prime correspond to the reference system and
$\mat{G}_0$ is the noninteracting Green's function. The symbol $\Tr$
denotes both a summation over bosonic Matsubara frequencies as well as
a summation over site indices. In order to be able to evaluate
$\Omega[\mat{\Sigma}]$ the self-energy $\mat{\Sigma}$ is approximated
by the self-energy of the reference system $\mat{\Sigma}^\prime$. In
practice this means that the functional $\Omega[\mat{\Sigma}]$ becomes
a function of single-particle parameters $\mat{x}$ of the reference
system $\hat{H}^\prime$ 
\begin{equation}
 \Omega(\mat{x}) = \Omega^\prime(\mat{x}) + \Tr\,\ln(-\mat{G}^{\prime}(\mat{x})) -
 \Tr\,\ln(-\mat{G}(\mat{x})) \;\mbox{.}
\label{eqn:om2}
\end{equation}
The stationary condition on $\Omega(\mat{x})$ now reads
\begin{equation}
 \PDF{\Omega (\mat{x})}{\mat{x}} = 0 \; \mbox{.}
 \label{eq:stat}
\end{equation}
The present formulation of VCA is not able to address the superfluid phase. Therefore our discussions will be restricted to Mott phase. A treatment of the superfluid phase would require an extension of the theory using Nambu formalism.

In VCA the reference system $\hat{H}^\prime$ is chosen to be a cluster
decomposition of the \psystem{}, which means that the total
lattice of $N$ sites is divided into decoupled clusters of size
$L$. Formally, the decomposition can be achieved by introducing a
superlattice, such that the physical lattice is obtained when a
cluster is attached to each site of the superlattice. The reference
system defined on one such cluster is solved using the band Lanczos
method.\cite{freund_roland_band_2000, aichhorn_variational_2006} It
can be carried out with open boundary conditions (obc), which is
generally done in literature,\cite{balzer_mott_2008} or with periodic
boundary conditions (pbc).\cite{potthoff_variational_2003} Both cases
are investigated in the next section. The Green's function of the
\psystem{} is obtained from the relation 
\begin{equation}
 \mat{G}^{-1}(\omega) = \mat{G}^{\prime\,-1}(\omega) - \mat{V} \; \mbox{,}
 \label{eq:num:gVCA}
\end{equation}
where
\begin{equation}
 \mat{V} \equiv -(\mu-\mu^\prime)\mathbbm{1} + (\mat{T}-\mat{T}^\prime)
 \label{eq:V}
\end{equation}
and $\mat{G}$, $\mat{G}^\prime$, and $\mat{V}$ are matrices in the site
indices. In the latter relation $\mat{T}$ and $\mat{T}^\prime$ are the
hopping matrices of the physical and reference system,
respectively. In order to evaluate the grand potential and the wave
vector and frequency resolved Green's function $G(\ve{k},\,\omega)$ of
the \psystem{} we apply the bosonic $\mat{Q}$-matrix
formalism.\cite{knap_spectral_2010} 
At first the bosonic $\mat{Q}$-matrix formalism yields the Green's
function of the \psystem{} in a mixed representation, partly in
real and partly in reciprocal space, $ \mat{G}(\tilde{\ve
  k},\,\omega)$. This representation is obtained by a partial Fourier
transform of \eq{eq:num:gVCA} from superlattice site indices to wave
vectors $\tilde{\ve k}$ of the first Brillouin zone of the
superlattice. $\mat{G}(\tilde{\ve{k}},\,\omega)$ is still a matrix in
cluster site indices. Second, we apply the Green's function
periodization proposed in Ref.~\onlinecite{snchal_spectral_2000} in
order to obtain the fully wave vector dependent Green's function
$G(\ve{k},\,\omega)$. 
From the Green's function $G(\ve{k},\,\omega)$ we are able to obtain the single-particle spectral function
\begin{equation}
 A(\ve{k},\,\omega)\equiv-\frac{1}{\pi} \mbox{Im} \, G(\ve{k},\,\omega)\;\mbox{,}
 \label{eq:spectralfunction}
\end{equation}
the density of states
\begin{equation}
 N(\omega)\equiv \frac{1}{N} \sum_{\ve{k}} A(\ve{k},\,\omega)
 \label{eq:dos}
\end{equation}
and the one-particle density matrix
\begin{equation}
 C(|\ve{r}_i-\ve{r}_j|) = \langle a_i^\dagger\,a_j \rangle\;\mbox{.}
 \label{eq:corr}
\end{equation}
The one-particle density matrix is the Fourier transform of the
momentum distribution $n(\ve{k})$, which can be evaluated in a
particularly accurate way by means of the $\mat{Q}$-matrix
formalism.\cite{knap_spectral_2010} 

The stationary point of $\Omega(\mat{x})$ is determined numerically by
varying some or all of the single-particle parameters of the reference
system, see \eq{eq:stat}. In the case of the BH model the
single-particle parameters are the hopping strength $t$ and the
chemical potential $\mu$. 
Due to the breaking of translation symmetry introduced by the
cluster partition,
the particle density evaluated as a trace of
$\mat{G}(\tilde{\ve{k}},\,\omega)$ differs at the boundary of
the cluster from the one inside the cluster. This is unfavorable as
the \psystem{}, which has pbc, should have a uniform particle
density. However, for reference systems with obc this problem could be
eased by adding another degree of freedom to the variational parameter
space, which allows for a different on-site energy at the
boundary of the cluster with respect to its bulk. Fortunately,
in VCA the reference system $\hat{H}^\prime$ can be extended by any
single-particle terms, as these terms do not affect the Legendre
transform of the Luttinger-Ward functional. It should be emphasized
that due to this fact adding single-particle parameters to the
variational parameter space does not affect the 
\psystem{}. However, additional, physically motivated variational
parameters improve the approximation of the self-energy $\mat{\Sigma}$
and therefore improved results for the grand potential of the 
\psystem{} and for other physical quantities are obtained. 
Clearly,
in the
\psystem{} the values of the parameters corresponding to the
additionally introduced single-particle terms are
zero, whereas in the reference system the values are determined by the
stationary condition on the grand potential. Due to these
considerations, the \psystem{} is not changed when introducing a
variable on-site energy at the boundary of the cluster.

\begin{figure}
        \centering
        \includegraphics[width=0.3\textwidth]{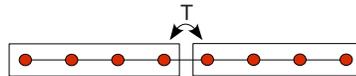}
        \caption{(Color online) A system of two clusters. The inter-cluster hopping $\mat{T}$ is treated perturbatively. }
        \label{fig:spe:clusterChainLatticePerturbation}
\end{figure}

For the
1D BH model the boundary of the cluster consists of the first
and the last cluster site. After adding the additional boundary
on-site energy the reference Hamiltonian of cluster $m$ is given by 
\begin{align}
 \hat{H}_m^\prime &= -t^\prime \sum_{\left\langle \alpha,\,\beta \right\rangle} b_\alpha^\dagger \, b_\beta + \frac{U^\prime}{2} \sum_\alpha \hat{n}_\alpha\left(\hat{n}_\alpha-1 \right) \nonumber \\
 &- \mu^\prime\,\sum_\alpha \hat{n}_\alpha + \delta^\prime(\hat{n}_1 + \hat{n}_L)\;\mbox{,}
 \label{eq:spe:HRefCluster}
\end{align}
where $\delta^\prime$ is the additionally introduced variational
parameter. In order to retain the chemical potential at approximately
the same level as it would be without the $\delta^\prime$ variation
the term $-\frac{2\delta^\prime}{L-2}\sum_{i=2}^{L-1} \hat{n}_i$ is
added to the Hamiltonian $\hat{H}_m^\prime$ as well. In VCA there is
no need to justify the
additional on-site
energy in the reference system, as this term is not included in
  the \psystem{}.
However, it is important to physically motivate the addition of
variational
parameters and, therefore, we show
below that the additional
on-site energy can be deduced from perturbation theory.
Consider two clusters as visualized in
\fig{fig:spe:clusterChainLatticePerturbation}, where the inter-cluster
hopping $\mat{T}$ from cluster $m=1$ to cluster $m=2$ is treated
perturbatively. 
The two cluster system is described by the Hamiltonian
\begin{equation}
 \hat{H} = \left( \begin{array}{cc}
 \hat{H}_1^\prime & \hat{T} \\
 \hat{T}^\dagger & \hat{H}_2^\prime
\end{array}
 \right) \;\mbox{.}
 \label{eq:spe:H2Cluster}
\end{equation}
The Schur decomposition of \eq{eq:spe:H2Cluster} yields
\begin{equation}
 \hat{H}_{2,\mbox{est}}^\prime = \hat{H}_2^\prime - \hat{T}^\dagger\, \hat{H}_1^{\prime\,-1}\,\hat{T} \;\mbox{.}
 \label{eq:spe:Hcpert}
\end{equation}
All parameters of the adjacent cluster are treated on mean field level
and thus we set $\hat{H}_1^\prime  \equiv \langle E\rangle$. Next we
have to calculate $\hat{T}^\dagger\, \hat{T}$, where
$\hat{T}=b_{1L}^{\dagger}\,b_{21}^{}+b_{21}^{\dagger}\,b_{1L}^{}$. The
notation $b_{m\alpha}$ was used, where $m$ denotes the cluster index
and $\alpha$ the site index within the cluster. With that we have 
\begin{align}
 \hat{T}^\dagger\, \hat{T} &= (b_{21}^{\dagger}\,b_{1L}^{}+b_{1L}^{\dagger}\,b_{21}^{})(b_{1L}^{\dagger}\,b_{21}^{}+b_{21}^{\dagger}\,b_{1L}^{}) \nonumber \\
 &= b_{21}^{\dagger}\,b_{1L}^{}\,b_{1L}^{\dagger}\,b_{21}^{}+b_{1L}^{\dagger}\,b_{21}^{}\,b_{21}^{\dagger}\,b_{1L}^{} \nonumber \\
 &= 2\,b_{21}^{\dagger}\,b_{21}^{}\,b_{1L}^{\dagger}\,b_{1L}^{} + b_{21}^{\dagger}\,b_{21}^{} + b_{1L}^{\dagger}\,b_{1L}^{} \nonumber \\
 &= \hat{n}_{21} (2\hat{n}_{1L} +1) + \hat{n}_{1L} \nonumber \\
 &= \hat{n}_{21} (2\langle{n}_{1L}\rangle +1) + \langle{n}_{1L}\rangle\;\mbox{.}
\end{align}
In the second step, we neglected the simultaneous hopping of two
particles from one cluster to the other and in the last step we
replaced the particle number operator of the adjacent cluster by a
mean field approximation. The constant energy shift
$\langle{n}_{1L}\rangle$ can be ignored and therefore
\eq{eq:spe:Hcpert} leads to 
\begin{equation}
 \hat{H}_{2,\mbox{est}}^\prime = \hat{H}_2^\prime - \frac{2\langle{n}_{1L}\rangle +1}{\langle E \rangle} \,\hat{n}_{21} \equiv   \hat{H}_2^\prime + \delta^\prime \,\hat{n}_{21} \;\mbox{,}
 \label{eq:spe:Hcpert2}
\end{equation}
where, in the second step, we replaced  the fraction, which is just an unknown constant, by $\delta^\prime$. Assuming that the \psystem{} has pbc and reiterating the above described procedure yields an extra term $\delta^\prime \hat{n}_{2L}$. Thus we obtain in total
\begin{equation}
 \hat{H}_{2,\mbox{est}}^\prime = \hat{H}_2^\prime + \delta^\prime (\hat{n}_{21} + \hat{n}_{2L})\;\mbox{.}
 \label{eq:spe:Hcpert3}
\end{equation}
In this way, we physically justified the additional on-site energy at the boundary of the cluster. In summary, we consider three possible variational parameters, namely the chemical potential $\mu$, the hopping strength $t$ and the additional on-site energy at the cluster boundary $\delta$.

\section{\label{sec:spectral}Spectral properties}
The first benchmark for VCA consists of a detailed investigation of the spectral properties of the one-dimensional BH model. In particular, we investigate the convergence properties of VCA for the phase boundary delimiting the first Mott lobe with respect to distinct sets of variational parameters, cluster sizes, and boundary conditions of the reference system. Moreover, we study and discuss the consequences of different variational parameters and boundary conditions of the reference system on the single-particle spectral functions and densities of states.

In the calculations, we use the following combinations of variational parameters $\mat{x}=\lbrace \mu \rbrace,\,\lbrace \mu,\, t \rbrace,\,\lbrace \mu,\,\delta \rbrace$ and $\lbrace \mu, t, \delta \rbrace$. 
It should be pointed out that the parameters of the references system are varied. Those of the \psystem{} are not modified.
We always consider the chemical potential $\mu$ as a variational parameter, since it has been shown that the chemical potential $\mu$ must be varied in order to obtain the correct particle density for the \psystem{}.\cite{koller_variational_2006, aichhorn_antiferromagnetic_2006} 
In general, CPT fails for bosonic systems, since the chemical potential of the reference systems leads to erroneous densities in both, the reference system and the 
\psystem{} accompanied by unphysical results, such as complex quasiparticle energies.

\begin{figure}
        \centering
        \includegraphics[width=0.48\textwidth]{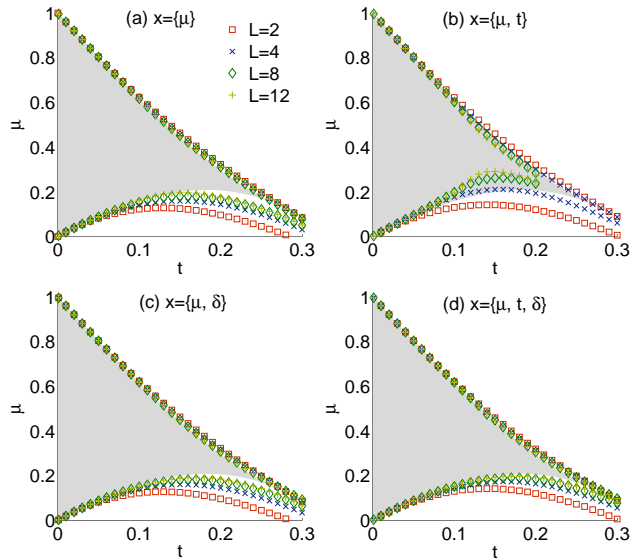}
        \caption{(Color online) First Mott lobe of the 1D BH model obtained for reference systems with obc and the variational parameter sets \fc{a} $\mat{x}=\lbrace \mu \rbrace$, \fc{b} $\mat{x}=\lbrace \mu,\, t \rbrace$, \fc{c} $\mat{x}=\lbrace \mu,\, \delta \rbrace$ and \fc{d} $\mat{x}=\lbrace \mu,\,t,\, \delta \rbrace$. The gray shaded area plotted in all subfigures indicates DMRG results obtained in Ref.~\onlinecite{khner_one-dimensional_2000}. }
        \label{fig:spe:bhphasediagram}
\end{figure}

Within the first Mott lobe, the particle density $n$ has to be one. After determining the stationary point of the grand potential $\Omega (\mat{x})$ with respect to the single-particle parameters $\mat{x}$ we always validate the particle density $n$, which is at $T=0$ given by
\begin{equation}
 n = -\frac{1}{N} \sum_{\ve k} \sum_{\lambda_m(\ve k)<0} a_m(\ve k) \;\mbox{,}
 \label{eq:spe:pd}
\end{equation}
where $\lambda_m(\ve k)$ are the poles of the Green's function and $a_m(\ve k)$ are their spectral weights. The phase boundaries for the first Mott lobe are shown in \fig{fig:spe:bhphasediagram} for reference systems with obc, the variational parameter sets $\mat{x}=\lbrace \mu \rbrace,\,\lbrace \mu,\, t \rbrace,\,\lbrace \mu,\,\delta \rbrace$, and $\lbrace \mu, t, \delta \rbrace$ and various cluster sizes $L$.
The gray shaded area shown in all four subfigures displays DMRG
results obtained from {T.~D.~K{\"u}hner} \textit{et al.} in
Ref.~\onlinecite{khner_one-dimensional_2000}. Additional work on the Mott to superfluid phase boundary can be found for instance in Refs.~\onlinecite{khner_phases_1998}, \onlinecite{Elstner_Dynamics_1999}, and \onlinecite{koller_variational_2006} and references therein. It can be observed that
all sets of variational parameters yield reasonable results apart from
the $\mat{x}=\lbrace \mu,\, t \rbrace$ variation. The best result is
achieved using the set $\mat{x}=\lbrace \mu,\,t,\, \delta \rbrace$ as
variational parameters. Particularly, the width of the phase diagram
is approximated very well even at larger hopping strength $t$, where
correlation effects are most pronounced, and the slope of the lobe tip
is obtained correctly. At the point shaped lobe tip
 a Berenzinskii-Kosterlitz-Thouless transition to a 
(quasi-long range ordered) superfluid phase occurs.\cite{fisher_boson_1989, khner_one-dimensional_2000, kosterlitz_ordering_1973} 
The quality of the calculated phase boundary can be quantified by $\chi$, the mean deviation of the VCA results from the DMRG data
\begin{equation}
 \chi = \frac{1}{M_p} \sum_i \left| p_i^{V} - p_i^{D} \right| \,\mbox{,}
 \label{eq:spe:chiQuality}
\end{equation}
where $p_i^{V}$ and $p_i^{D}$ are corresponding phase boundary points calculated by means of VCA and DMRG, respectively, and $M_p$ is the number of phase boundary points, which contribute to the sum. The quality $\chi$ is stated in \tab{tab:spe:quality} for several sets of variational parameters and cluster sizes.
\begin{table}
\caption{Quality $\chi/10^{-3}$ of the phase boundary for reference systems with obc and pbc shown in \figg{fig:spe:bhphasediagram}{fig:spe:bhphasediagrampbc}, respectively. The quality $\chi$ is evaluated using \eq{eq:spe:chiQuality}. }
        \label{tab:spe:quality}
        \centering
        \begin{tabular}{rcl}
    $L$ & \ldots & number of cluster sites  \\
    $\mu,\, t,\,\delta$ & \ldots & variational parameters \\

        \end{tabular} \newline
        \begin{tabular}{|c||c|c|r @ {.} l|r @ {.} l|r @ {.} l|r|}
                \hline
                & \multicolumn{6}{c|}{\emph{obc}} & \multicolumn{3}{c|}{\emph{pbc}} \\ \hline
                \emph{$L$} & \emph{$\mu$} & \emph{$\mu,\, t$} &  \multicolumn{2}{c|}{\emph{$\mu,\, \delta$}} & \multicolumn{2}{c|}{\emph{$\mu,\, t,\,\delta$}}  & \multicolumn{2}{c|}{\emph{$\mu$}} & \emph{$\mu,\, t$} \\
                \hline
                \hline

        2 & 40.1 & 32.7 & 40&1 & {\;\;}32&7& 22&3&32.7\\ \hline
        4 & 23.8 & 10.5 & 22&7 & 16&5& 16&0&16.4\\ \hline
        8 & 14.3 & 21.1 & 12&1 & 8&6 &9&2&5.8\\ \hline
        12 & 11.5 & 40.1 & 8&7 & 6&1 &8&0&4.4\\ \hline
        \end{tabular}
\end{table}
From that it can be seen as well that for reference systems with obc
the $\mat{x}=\lbrace \mu,\,t,\, \delta \rbrace$ variation yields the
best approximation for the phase boundary, as compared with DMRG.

In conventional variational methods, such as Hartree Fock,
the ``best'' among  a set of solutions (at given $\mu$ and $T$) is chosen according to the
principle of minimum grand potential. 
This criterion cannot be applied in our
case, since there is no such minimum principle in VCA. 
In addition, when including an additional variational parameter
such as  $\delta$, 
there is no reason why 
the grand potential $\Omega$ should display a saddle point at $\delta=0$
since $\Omega$ is not an even function of $\delta$.

\begin{figure}
        \centering
        \includegraphics[width=0.48\textwidth]{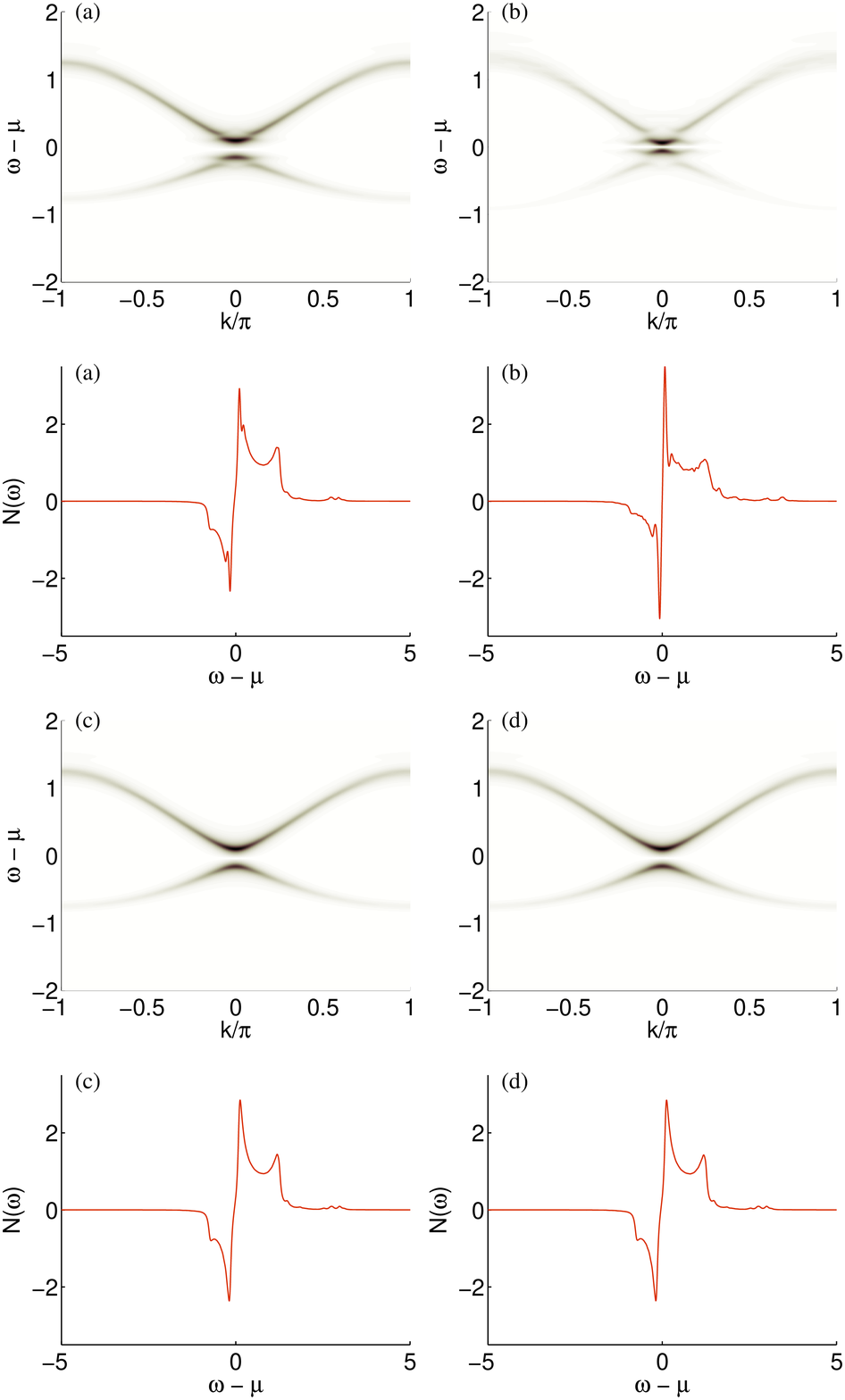}
        \caption{(Color online) Spectral function $A(\ve{k},\,\omega)$ and density of states $N(\omega)$ for $t=0.15$, $\mu=0.35$, reference systems of $L=12$ sites with obc and the following sets of variational parameters \fc{a} $\mat{x}=\lbrace \mu \rbrace$, \fc{b} $\mat{x}=\lbrace \mu,\, t \rbrace$, \fc{c} $\mat{x}=\lbrace \mu,\, \delta \rbrace$, and \fc{d} $\mat{x}=\lbrace \mu,\,t,\, \delta \rbrace$. }
        \label{fig:spe:bhsf}
\end{figure}

The single-particle spectral function $A(\ve{k},\,\omega)$ has been evaluated for $t=0.15$ and $\mu=0.35$, which corresponds to a point right in the middle of the first Mott lobe. It is shown in \fig{fig:spe:bhsf} along with the corresponding density of states $N(\omega)$ for reference systems with obc and $L=12$ sites, and distinct sets of variational parameters.
For the numerical evaluation we used an artificial broadening
$\eta=0.05$. Close to the main gap at $\omega-\mu=0$ the spectral
functions obtained from the $\lbrace \mu \rbrace$ and $\lbrace
\mu,\,t\rbrace$ variation are not smooth but exhibit spurious
gaps,\cite{koller_variational_2006} see
\figcc{fig:spe:bhsf}{a}{b}. However, as soon as the variation in
$\delta$ is considered the spectral function becomes smooth, see
\figcc{fig:spe:bhsf}{c}{d}. This also manifests in the density of
states. Interestingly, in contrast to the results in one dimension,
there are no visible spurious gaps in the spectral functions of the
two-dimensional BH model when only the variation of the chemical
potential $\mat{x}=\lbrace \mu \rbrace$ is
considered.\cite{knap_spectral_2010} This can be explained by the fact
that in two-dimensions most of the cluster sites are actually
boundary sites as well.

\begin{figure}
        \centering
        \includegraphics[width=0.48\textwidth]{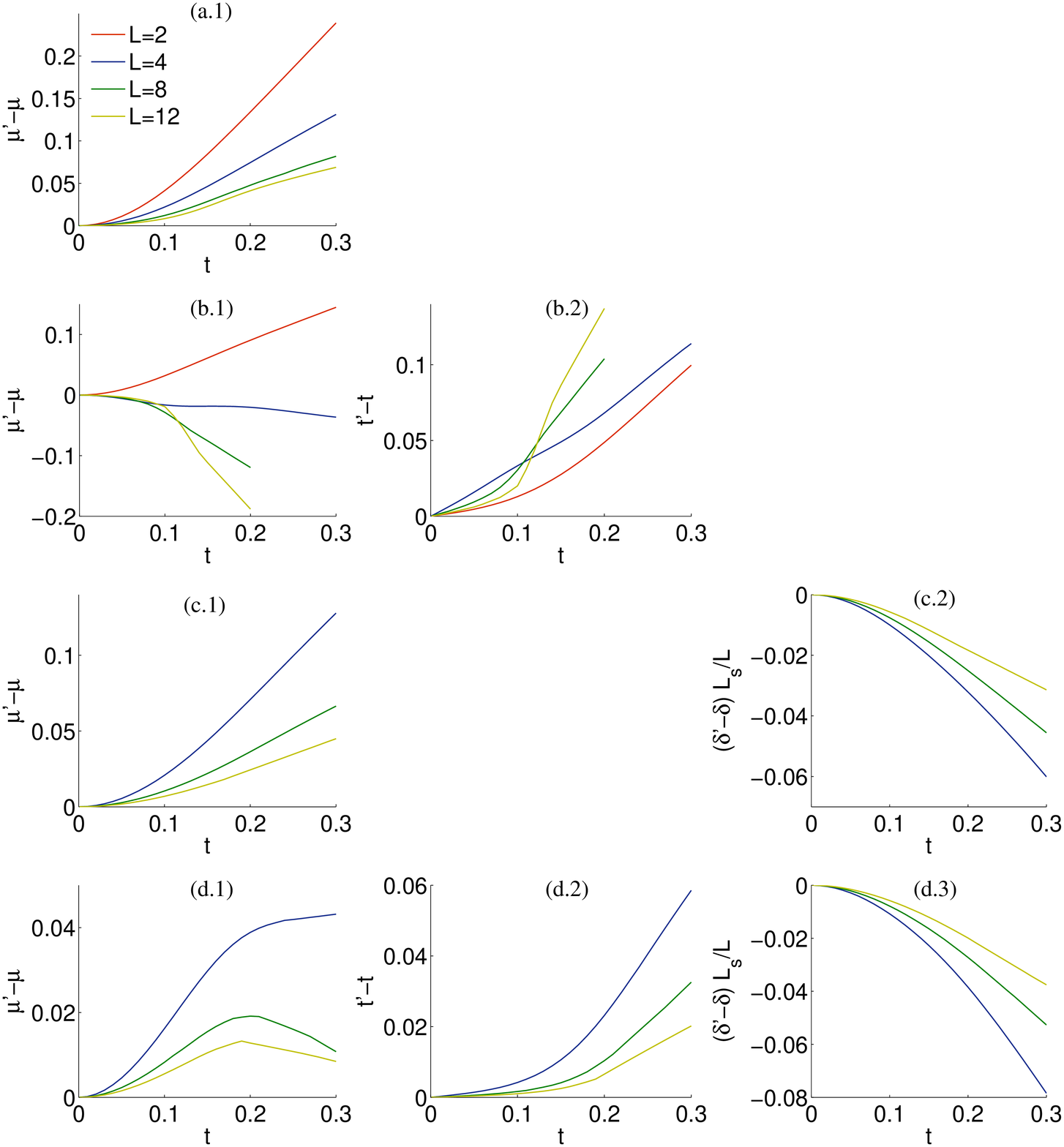}
        \caption{(Color online) Difference between the parameters of
          the reference system $\hat{H}^\prime$ at the stationary
          point $\lbrace \mu^\prime,\,t^\prime,\,\delta^\prime\rbrace$
          and the parameters $\lbrace \mu,\,t,\, \delta=0 \rbrace$ of
          the \psystem{} $\hat{H}$. The corresponding variational
          parameter sets are \fc{a.$\ast$} $\mat{x}=\lbrace \mu
          \rbrace$, \fc{b.$\ast$} $\mat{x}=\lbrace \mu,\, t \rbrace$,
          \fc{c.$\ast$} $\mat{x}=\lbrace \mu,\, \delta \rbrace$, and
          \fc{d.$\ast$} $\mat{x}=\lbrace \mu,\,t,\, \delta \rbrace$.} 
        \label{fig:spe:bhdev}
\end{figure}

\begin{figure}
        \centering
        \includegraphics[width=0.48\textwidth]{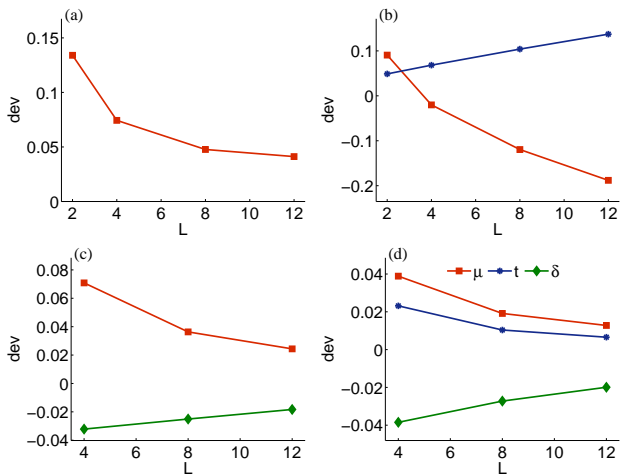}
        \caption{(Color online) Difference between the parameters of the reference system $\hat{H}^\prime$ at the stationary point $\lbrace \mu^\prime,\,t^\prime,\,\delta^\prime\rbrace$ and the parameters $\lbrace \mu,\,t=0.2,\, \delta=0 \rbrace$ of the \psystem{} $\hat{H}$ in dependence of the cluster size $L$. The corresponding variational parameter sets are \fc{a} $\mat{x}=\lbrace \mu \rbrace$, \fc{b} $\mat{x}=\lbrace \mu,\, t \rbrace$, \fc{c} $\mat{x}=\lbrace \mu,\, \delta \rbrace$, and \fc{d} $\mat{x}=\lbrace \mu,\,t,\, \delta \rbrace$.}
        \label{fig:spe:bhdevExtract}
\end{figure}

The $\lbrace \mu,\,t\rbrace$ variation has an odd behavior, see \figc{fig:spe:bhphasediagram}{b}. Naively one would expect that the results should improve for larger clusters and the more variational parameters are used. However, this seems not to be valid for the variational parameter set $\mat{x}=\lbrace \mu,\,t\rbrace$. Therefore, this case has to be studied in more detail.
An important aspect is, that the larger the cluster the better should
be the approximation to the \psystem{}. 
Therefore, one
would expect that the deviations of the parameters of the reference
system $\hat{H}^\prime$ at the stationary point $\lbrace
\mu^\prime,\,t^\prime,\,\delta^\prime\rbrace$ from  the parameters $\lbrace \mu,\,t,\, \delta=0 \rbrace$ of the \psystem{} $\hat{H}$ should decrease with increasing cluster size. 
At the same time, one would also expect the VCA results to come
closer to the exact ones.
The parameter deviations are shown in \fig{fig:spe:bhdev} for various values of the hopping strength $t$ and in \fig{fig:spe:bhdevExtract} for $t=0.2$ as a function of the cluster size $L$. Please notice that as a result of the properties of the Mott phase the difference between $\mu$ and $\mu^\prime$ is independent of the actual value of $\mu$ provided the values of $\mu$ are restricted to the same Mott lobe.
For the variational parameters $\mat{x}=\lbrace \mu \rbrace$, $\lbrace \mu,\, \delta \rbrace$ and $\lbrace \mu,\,t,\, \delta \rbrace$ the deviations decrease steadily with increasing cluster size. The variational parameter $\delta$ is somewhat an exception as the average value of $\delta$ on the cluster is $\delta L_s/L$, where $L_s$ denotes the number of boundary sites of the cluster, \ie, $L_s=2$ for one-dimensional lattices. Correspondingly, the scaled value of $\delta$ is plotted in \figg{fig:spe:bhdev}{fig:spe:bhdevExtract}.
Clearly, the results for the two-site cluster are the same for the
$\lbrace \mu \rbrace$ and $\lbrace \mu,\, \delta \rbrace$ variation,
and for the $\lbrace \mu,\,t \rbrace$ and $\lbrace \mu,\,t,\, \delta
\rbrace$ variation, respectively. Thus the deviations for that cluster
size are not shown in \figcc{fig:spe:bhdev}{c.$\ast$}{d.$\ast$}. In
comparison to the deviations for all other parameter sets the
deviations for $\mat{x}=\lbrace \mu,\,t \rbrace$ depicted in
\figc{fig:spe:bhdev}{b.$\ast$} show a completely different
behavior. In particular, they do not decrease with increasing
cluster size and furthermore they cross each other. This behavior
suggests that the results might be incorrect.

\begin{figure}
        \centering
        \includegraphics[width=0.48\textwidth]{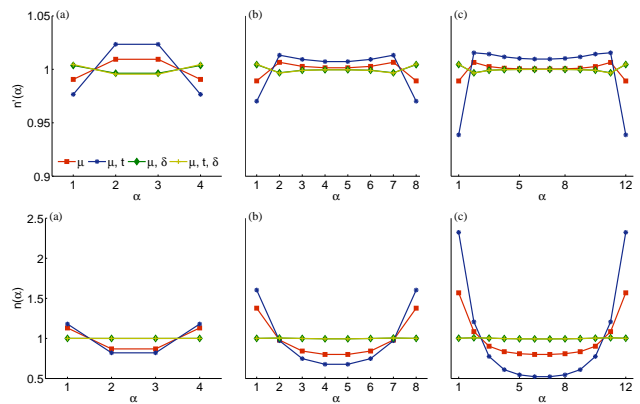}
        \caption{(Color online) Particle density $n^\prime(\alpha)$, first row, and $n(\alpha)$, second row, for reference systems with obc and size \fc{a} $L=4$, \fc{b} $L=8$, and \fc{c} $L=12$. The parameters for the \psystem{} are $t=0.15$ and $\mu=0.35$. The legend refers to the distinct sets of variational parameters. }
        \label{fig:spe:bhpartdensclu}
\end{figure}

The additionally introduced variational parameter $\delta$ drastically
improves the results and has significant influence on the convergence
properties.   We suggested above that $\delta$ allows
for a more uniform distribution of the particle density
within the cluster. To demonstrate that this is indeed the case we
determine both the particle density $n^\prime(\alpha)$ obtained from
the cluster Green's function $\mat{G}^\prime(\omega)$ evaluated at the
stationary point of the grand potential as well as the particle
density $n(\alpha)$ obtained from the VCA Green's function
$\mat{G}(\tilde{\ve k},\,\omega)$ of the \psystem{}. The latter
is given partly in real and partly in reciprocal space. It is
important to note that at this point the Green's function
$\mat{G}(\tilde{\ve k},\,\omega)$ has not yet been periodized, and
that the index $\alpha$ is a cluster site index. Thus, it ranges from
$1$ to $L$. The particle density $n(\alpha)$ is evaluated by
calculating the trace of the Green's function $\mat{G}(\tilde{\ve
  k},\,\omega)$, which reduces at $T=0$ to a sum over the residues of
the Green's function corresponding to poles with negative energy and a
sum over the wave vectors $\tilde{\ve k}$. The results for the
particle densities $n^\prime(\alpha)$ and $n(\alpha)$, respectively,
for reference systems with obc and of different size are shown in
\fig{fig:spe:bhpartdensclu}. 
As one can see from the figure,
the particle distribution $n^\prime(\alpha)$, first row, becomes
flatter when $\delta$ is considered as variational parameter and the
deviations from density one shrink with increasing cluster size,
\ie, from the left to the right panel. The only exception is the
$\mat{x}=\lbrace \mu,\,t \rbrace$ variation, where deviations from one
increase for larger clusters. 
However, a uniform
particle density $n(\alpha)$ 
in the \psystem{}
is even more important. In VCA the lattice of the
\psystem{}, which has pbc and thus a uniform particle
density, is decomposed into clusters of size $L$. This breaks the
translational invariance of the \psystem{} and hence a
periodization prescription has to be applied ``by hand'' to the Green's function,
such that it obeys the translational invariance of the
physical lattice.
 The periodized Green's function $G(\ve k, \,\omega)$ depends
only on one wave vector $\ve k$ of the first Brillouin zone of the
physical lattice and therefore yields, per construction, a uniform
particle density of the physical lattice. 
Nevertheless, it would be desirable to obtain a particle density
$n(\alpha)$ which is as flat as possible {\em even without Green's function
  periodization}.
From \fig{fig:spe:bhpartdensclu}, second row, it can be observed that the particle density $n(\alpha)$ varies significantly with cluster lattice sites $\alpha$ when using the variational parameter sets $\mat{x}=\lbrace \mu \rbrace$ and $\lbrace \mu,\,t \rbrace$. The large deviations of $n(\alpha)$ from $1$ for these parameter sets seem to be related to the spurious gaps observed in the corresponding spectral functions of \figcc{fig:spe:bhsf}{a}{b}. However, as soon as $\delta$ is introduced as variational parameter $n(\alpha)$ is indeed absolutely flat.

\begin{figure}
        \centering
        \includegraphics[width=0.48\textwidth]{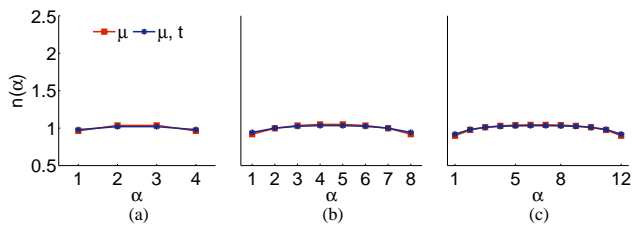}
        \caption{(Color online) Particle density $n(\alpha)$ obtained from $\mat{G}(\tilde{\ve k},\,\omega)$ for the parameters $t=0.15$ and $\mu=0.35$, and reference systems with pbc. The clusters are of size \fc{a} $L=4$, \fc{b} $L=8$, and \fc{c} $L=12$. The legend refers to the considered sets of variational parameters. }
        \label{fig:spe:bhpartdensclupbc}
\end{figure}

Alternatively to obc we consider pbc for the reference Hamiltonian. The advantage of pbc is that the  particle density $n^\prime(\alpha)$ within  the cluster is uniform. However, this does not necessarily imply that the particle density $n(\alpha)$ obtained from $\mat{G}(\tilde{\ve k},\,\omega)$ is flat as well,
since the additional hopping term between the boundary points of the cluster have to be subtracted again in  VCA via the matrix $\mat V$, 
which again breaks the translational symmetry.
For reference systems with pbc the variation in $\delta$ appears
  less meaningful. 
Therefore, we consider only the variational parameter sets
$\mat{x}=\lbrace \mu \rbrace$ and $\lbrace \mu,\,t \rbrace$. Due to
the pbc there are additional contributions of the hopping over the
cluster border which are not present in the full system. This
contribution has to be subtracted using extra terms in the matrix
$\mat{V}_{pbc}$, see \eq{eq:V}. As we did throughout this paper, we
consider here a one-dimensional lattice to deduce the matrix
$\mat{V}_{pbc}$. However, the results can be readily extended to
higher dimensions. The hopping matrices of the full system $\mat{T}$
and of the reference system $\mat{T}^\prime_{pbc}$ are given by 
\begin{equation}
 (\mat{T})_{ij} = -t\,(\delta_{ij+1} + \delta_{i+1j})
\end{equation}
and
\begin{equation}
 (\mat{T}^\prime_{pbc})_{ij} = (\mat{T}^\prime_{pbc})_{m\alpha n\beta} = -t^\prime \delta_{mn} (\delta_{\alpha \beta+1} + \delta_{\alpha+1 \beta}) \;\mbox{,}
 \label{eq:spe:Tprimepbc}
\end{equation}
where $\lbrace i,\,j \rbrace$ are site indices in the \psystem{}, $\lbrace m,\,n \rbrace $ label the clusters and $\lbrace \alpha, \beta \rbrace $ are site indices in the cluster. We use pbc in the indices $\lbrace i,\,j \rbrace $  and $\lbrace \alpha, \beta \rbrace$, \ie, $\lbrace i,\,j\rbrace =N+1 = 1$ and $\lbrace \alpha,\,\beta\rbrace =L+1 = 1$. After a partial Fourier transform from superlattice site indices $m$ to wave vectors $\tilde{\ve{k}}$ of the first Brillouin zone of the superlattice, one obtains
\begin{align}
(\mat{T}-\mat{T}^\prime_{pbc})_{\alpha\beta}(\tilde{\ve{k}}) &= -(t-t^\prime) (\delta_{\alpha \beta+1} + \delta_{\alpha+1 \beta})  \nonumber \\
 &+(t - t\,e^{-i\,\tilde{\ve{k}}}) \delta_{\alpha 1}\delta_{\beta L} \nonumber \\
 &+ (t - t\,e^{i\,\tilde{\ve{k}}}) \delta_{\alpha L}\delta_{\beta 1} \;\mbox{.}
 \label{eq:spe:Tdiffkspace}
\end{align}
In contrast to obc this matrix has an extra contribution of $-(t-t^\prime)+t = t^\prime$ at its top right and bottom left corner. The whole matrix $\mat{V}_{pbc}(\tilde{\ve{k}})$ contains the variation of the chemical potential as well and hence we have
\begin{align}
(\mat{V}_{pbc})_{\alpha\beta}(\tilde{\ve{k}}) &=  -(t-t^\prime) (\delta_{\alpha \beta+1} + \delta_{\alpha+1 \beta}) - (\mu - \mu^\prime) \delta_{\alpha \beta} \nonumber \\
 &+(t - t\,e^{-i\,\tilde{\ve{k}}}) \delta_{\alpha 1}\delta_{\beta L} \nonumber \\
 &+ (t - t\,e^{i\,\tilde{\ve{k}}}) \delta_{\alpha L}\delta_{\beta 1} \;\mbox{.}
 \label{eq:spe:VpbcKSpace}
\end{align}
\begin{figure}
        \centering
        \includegraphics[width=0.48\textwidth]{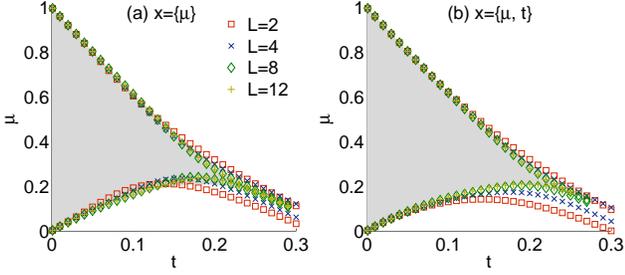}
        \caption{(Color online) Phase boundaries of the first Mott lobe of the 1D BH model. The reference system consists of clusters with pbc. The variational parameters  \fc{a} $\mat{x}=\lbrace \mu \rbrace$ and \fc{b} $\mat{x}=\lbrace \mu,\, t \rbrace$ are used. The gray shaded area shows the DMRG results from Ref.~\onlinecite{khner_one-dimensional_2000}.}
        \label{fig:spe:bhphasediagrampbc}
\end{figure}
\begin{figure}
        \centering
        \includegraphics[width=0.48\textwidth]{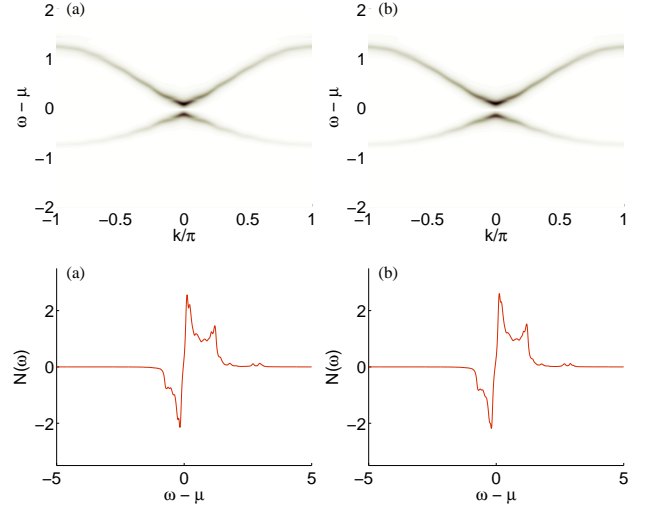}
        \caption{(Color online) Spectral function $A(\ve{k},\omega)$ and density of states $N(\omega)$ for a 12 site cluster with pbc as reference system. The parameters of the \psystem{} are $t=0.15$ and $\mu=0.35$. The corresponding variational parameter sets are \fc{a} $\mat{x}=\lbrace \mu \rbrace$ and \fc{b} $\mat{x}=\lbrace \mu,\, t \rbrace$.}
        \label{fig:spe:bhsfpbc}
\end{figure}
When treating clusters with pbc by means of VCA the replacement
of the matrix $\mat{V}$ by $\mat{V}_{pbc}$ is the only new aspect
which has to be considered apart from the fact that the solution of
the cluster itself is different due to the pbc. The first Mott lobe of
the phase diagram for the variational parameter sets
$\mat{x}=\left\lbrace \mu \right\rbrace$ and $\mat{x}=\left\lbrace
  \mu,\,t \right\rbrace$,
with a reference Hamiltonian with pbc are shown in \fig{fig:spe:bhphasediagrampbc}.
\begin{figure}
        \centering
        \includegraphics[width=0.48\textwidth]{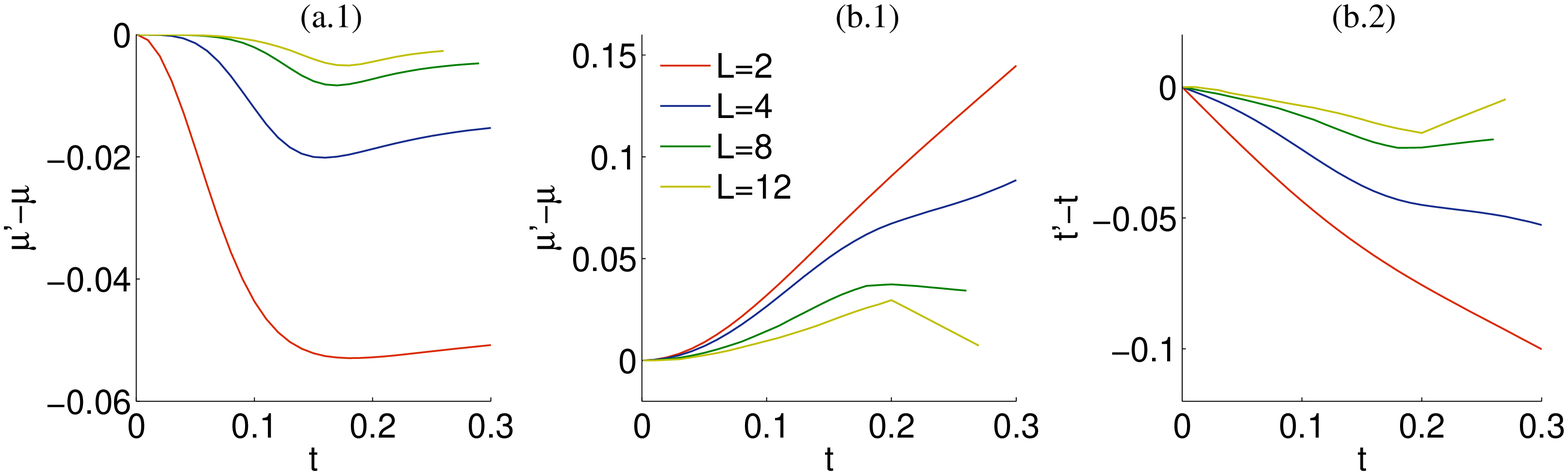}
        \caption{(Color online) Difference between the variational parameters at the stationary point of the grand potential and the parameters of the \psystem{} for reference systems with pbc. The variational parameter sets \fc{a.$\ast$} $\mat{x}=\lbrace \mu \rbrace$ and \fc{b.$\ast$} $\mat{x}=\lbrace \mu,\, t \rbrace$ are used.}
        \label{fig:spe:bhdeviationspbc}
\end{figure}
Both Mott lobes coincide reasonably with the DMRG data. The phase boundary obtained from the $\lbrace \mu,\, t \rbrace$ variation yields a very good agreement and the problems which occurred for this parameter set for reference systems with obc are not present anymore. The quality $\chi$ of the phase boundary calculated according to \eq{eq:spe:chiQuality} is stated in \tab{tab:spe:quality}. From that it can be seen that the $\lbrace \mu,\, t \rbrace$ variation with pbc is comparable to the $\lbrace \mu,\, t,\,\delta \rbrace$ variation with obc. The spectral function $A(\ve{k}, \omega)$ and the corresponding density of states $N(\omega)$ is shown in \fig{fig:spe:bhsfpbc} for a 12 site cluster and the parameters $t=0.15$ and $\mu=0.35$. 
\begin{figure}
        \centering
        \includegraphics[width=0.48\textwidth]{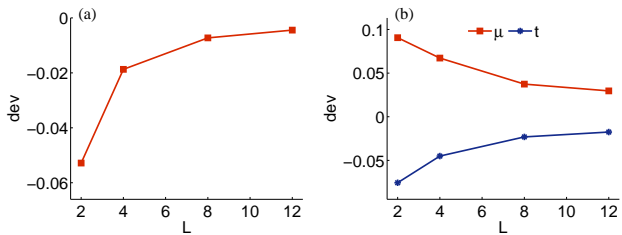}
        \caption{(Color online) Difference between the variational parameters at the stationary point of the grand potential and the parameters of the \psystem{} for reference systems with pbc for the hopping strength $t=0.2$ in dependence of the cluster size $L$. The variational parameter sets are \fc{a} $\mat{x}=\lbrace \mu \rbrace$ and \fc{b} $\mat{x}=\lbrace \mu,\, t \rbrace$.}
        \label{fig:spe:bhdeviationspbcExtract}
\end{figure}
In the case of pbc the spectral function and the density of states are not as smooth as in the case of obc and $\mat{x}=\lbrace \mu,\,t,\,\delta \rbrace$ variation but overall they exhibit the important features.
The reason for this is that for pbc there are more conservation laws and therefore less states, so the states are less dense.
The deviation of the variational parameters with respect to the system parameters is as demanded shrinking for increasing cluster size, see \figg{fig:spe:bhdeviationspbc}{fig:spe:bhdeviationspbcExtract}.
Next we investigate the particle density $n(\alpha)$ obtained from the not yet periodized Green's function $\mat{G}(\tilde{\ve k},\,\omega)$. Results are shown in \fig{fig:spe:bhpartdensclupbc}.
Interestingly, for a reference systems with pbc the density is less uniform than for a reference system with obc 
which has an additional boundary potential described  by the variational parameter $\delta$.

\begin{figure}
        \centering
        \includegraphics[width=0.48\textwidth]{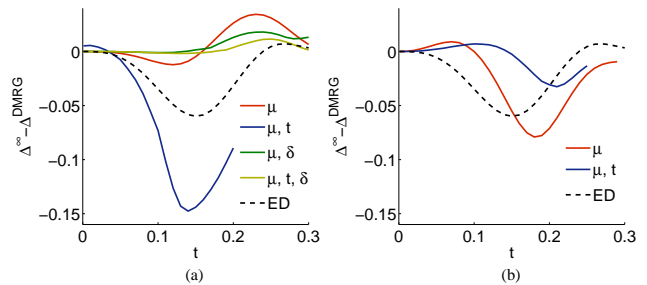}
        \caption{(Color online) Deviations between the extrapolated gaps $\Delta^\infty$ (VCA) and the gap $\Delta^{DMRG}$ (DMRG), where \fc{a} shows the deviations for reference systems with obc and \fc{b} for reference systems with pbc. The dashed lines correspond to exact diagonalization results for the gap extrapolated to infinitely large systems.}
        \label{fig:spe:gapDev}
\end{figure}

Next we perform finite size scaling for the energy gap
  $\Delta^L$. Assuming a $1/L$ dependence 
($L$ is the cluster size) 
we estimate the
    infinite-system gap $\Delta^\infty$. 
Notice that, in principle, 
VCA provides the gap for an infinite system. However, due to the
approximation, the gap displays a
dependence on the cluster size $L$, which should converge toward the exact value for $L\to \infty$.
In \fig{fig:spe:gapDev}  we show 
the deviations of the gap $\Delta^\infty$  from the gap obtained from DMRG $\Delta^{DMRG}$.
One  can observe that  in this case the best results are obtained for the parameter set $ \mat{x}=\lbrace \mu,\, t,\,\delta \rbrace $ and obc for the reference system. The scaled results for reference systems with pbc are not as good as the scaled results for reference systems with obc. At the first sight it might be counterintuitive that the agreement between the VCA and DMRG results is good close to the lobe tip. Yet, it should be noticed that the gap shrinks rapidly with increasing hopping strength and that we show in \fig{fig:spe:gapDev} the absolute error rather than the relative one.

We also compared the extrapolated VCA results with extrapolated exact diagonalization (ED) results, where we used systems with pbc of up to $12$ lattice sites. 
The best VCA results (for $\mat{x}=\lbrace \mu,\, \delta,\, t \rbrace$ and a reference system with obc) are significantly superior 
to the best results obtained by extrapolating the bare ED data. The discrepancy is particularly pronounced deep in the Mott lobe for
$t\approx 0.15$.

In all spectral functions, regardless of the boundary conditions, we observe additionally to the two pronounced cosinelike shaped bands centered around $\omega - \mu = 0$ other bands with little spectral weight located at higher energies.\cite{additionalbands} The intensity and width of these bands increases with increasing hopping strength $t$. In order to understand the additional bands we apply first-order perturbation theory on the ground state, where we consider the hopping term of the BH Hamiltonian, see \eq{eq:bhm}, as perturbation, \ie, $\hat{H}_1=-t \sum_{\left\langle i,\,j \right\rangle} b_i^\dagger \, b_j $. The ground state $\ket{\psi_0^{(0)}}$ of the unperturbed system with particle density $n$ is given by
\[
 \ket{\psi_0^{(0)}} = \bigotimes_{\nu=1}^N \ket{n}_\nu \;\text{,}
\]
which is a tensor product of the individual single-site bare states $\ket{n}$. The energy of the bare states is $E_{\ket{n}} = U \, n(n-1)/2 - \mu\,n$. The first-order correction of the ground state describing quantum fluctuations is of the form
\begin{equation}
 \ket{\Delta\psi_0^{(1)}} = \frac{t}{\Delta E} \ket{n+1}_l \otimes \ket{n-1}_{l^\prime} \bigotimes_{ \genfrac{}{}{0pt}{}{\nu = 1}{\nu \neq l,l'} }^N \ket{n}_\nu \;\text{,}
 \label{eq:gsPert}
\end{equation}
where $l$ and $l^\prime$ are nearest neighbors and $\Delta E = E_{\ket{n+1}} +  E_{\ket{n-1}} -2\, E_{\ket{n}}$. The first-order correction $\ket{\Delta\psi_0^{(1)}}$ is proportional to the hopping strength $t$, which reflects the fact that the intensity of the additional bands increases with increasing hopping strength.

\begin{figure}
        \centering
        \includegraphics[width=0.48\textwidth]{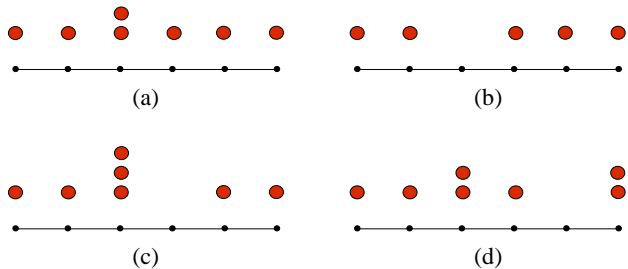}
        \caption{(Color online) Small black dots indicate lattice sites and big red dots particles. \fc{a} single-particle term of the Green's function and \fc{b} single-hole term of the Green's function. In both situations the additional particle/hole can move freely on the lattice. \fc{c} and \fc{d}, possible higher order excitations due to quantum fluctuations in the ground state of the single-particle term of the Green's function. }
        \label{fig:spe:bh1DadBands2}
\end{figure}

\begin{figure}
        \centering
        \includegraphics[width=0.48\textwidth]{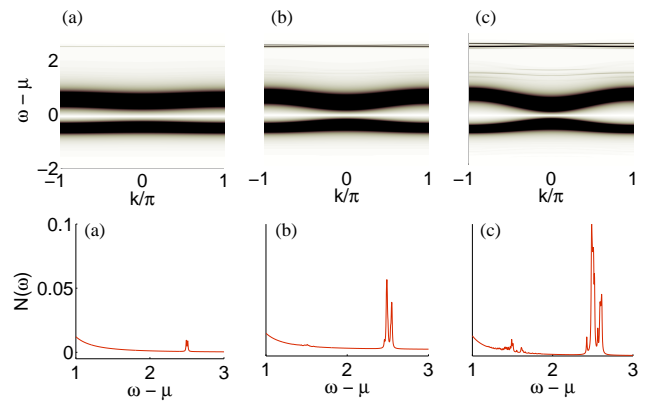}
        \caption{(Color online) Spectral functions, first row, and density of states, second row, for \fc{a} $t=0.01,\,\mu=0.5$, \fc{b} $t=0.03,\,\mu=0.5$ and \fc{c} $t=0.05,\,\mu=0.5$. For the visualization we used a threshold of $0.1$. The reference system consists of an $L=12$ site cluster with obc. }
        \label{fig:spe:bh1DadBands}
\end{figure}

In the first Mott lobe each lattice site is occupied by a single particle, provided quantum fluctuations are neglected. Hence, the additional particle in the particle part
of the Green's function can move freely, see \figc{fig:spe:bh1DadBands2}{a}.
This yields the pronounced upper band of the spectral function located at $\omega - \mu \approx 0.5$. The cosinelike shape of the band is reminiscent of the dispersion relation of free particles propagating on a lattice. Equivalently, the hole, introduced in the hole part of the Green's function, gives rise to the pronounced lower band, see \figc{fig:spe:bh1DadBands2}{b}.
Considering quantum fluctuations in the ground state, see \eq{eq:gsPert}, there are three energetically distinct excitations possible, namely, excitations from $\ket{n+1}$ to $\ket{n+2}$, from $\ket{n}$ to $\ket{n+1}$ and from $\ket{n-1}$ to $\ket{n}$. Next we evaluate the location of the bands arising from these excitations. The first investigated situation corresponding to the excitation from $\ket{n+1}$ to $\ket{n+2}$ yields for the location of the band
\begin{align}
 \tilde{\omega}_p^1 &= E_{\ket{n+2}} +  E_{\ket{n-1}} +(N-2)\, E_{\ket{n}} - N \, E_{\ket{n}} \nonumber\\
 &\stackrel{n=1}{=} -\mu+3U\;\mbox{,}
 \label{eq:pertBand1}
\end{align}
where we used $n=1$ for the first Mott lobe in the second step. This situation, where one lattice site is occupied by three particles, is sketched in \figc{fig:spe:bh1DadBands2}{c}. The second situation corresponding to the excitation from $\ket{n}$ to $\ket{n+1}$ results in
\begin{align}
 \tilde{\omega}_p^2 &= 2\,E_{\ket{n+1}} +  E_{\ket{n-1}}+(N-3)\, E_{\ket{n}} - N \, E_{\ket{n}} \nonumber \\
 &\stackrel{n=1}{=} -\mu+2U
 \label{eq:pertBand2}
\end{align}
and is sketched in \figc{fig:spe:bh1DadBands2}{d}. The third situation (excitation from $\ket{n-1}$ to $\ket{n}$) corresponds to \figc{fig:spe:bh1DadBands2}{a} and therefore contributes to the pronounced cosinelike shaped band located at $\omega-\mu \approx 0.5$. In the following, we compare the perturbative results for the additional bands with the VCA results. Particularly, we evaluate spectral functions and densities of states for the variational parameters $ \mat{x}=\lbrace \mu,\,t,\,\delta \rbrace $, the chemical potential $\mu = 0.5$ and small hopping strength $t=\left\lbrace 0.01, 0.03, 0.05\right\rbrace$, see \fig{fig:spe:bh1DadBands}.
For small hopping strength $t$ the perturbative treatment is well suited. However, the quantum fluctuations in the ground state, and thus the spectral weight of the additional bands as well, are small, see \eq{eq:gsPert}. In order to uncover the additional bands in the spectral functions we introduce a threshold for the spectral weight, \ie, the maximum value of the spectral function is restricted to this threshold and all values larger than or equal to this threshold are plotted in black color in the figures. This unveils bands with very low intensity. In all spectral functions shown in \fig{fig:spe:bh1DadBands} an additional band located at $\omega_p^1 \approx 2.5$ can be observed. Furthermore a band located at $\omega_p^2 \approx 1.5$ is visible in \figcc{fig:spe:bh1DadBands}{b}{c}. These bands match perfectly well with the perturbative results $\tilde{\omega}_p^1$ and $\tilde{\omega}_p^2$ of \eqq{eq:pertBand1}{eq:pertBand2}, which are given by $\tilde{\omega}_p^1 = 2.5$ and $\tilde{\omega}_p^2 = 1.5$, respectively, where we employed a chemical potential $\mu=0.5$ and used $U$ as unit of energy.

\section{\label{sec:static} Static Properties}
\begin{figure}
        \centering
        \includegraphics[width=0.48\textwidth]{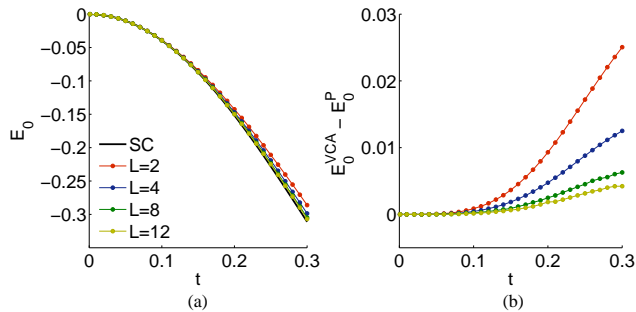}
        \caption{ (Color online) Comparison of the ground state energy $E_0$ for hopping parameters corresponding to the first Mott lobe obtained by means of VCA (dots connected by lines) and strong-coupling perturbation theory (solid line). The lines connecting the VCA results are guidance for the eyes. \fc{a} directly compares the ground state energy obtained from the two approaches and \fc{b} shows their difference. }
        \label{fig:energy}
\end{figure}

In this section, we benchmark the VCA results using the ground
state properties of the BH model. In particular we investigate the
ground state energy $E_0$ and the one-particle density matrix
$C(|\ve{r_i}-\ve{r_j}|) \equiv \langle a_i^\dagger\,a_j\rangle$, which
both have been evaluated in
Ref.~\onlinecite{damski_mott-insulator_2006} by means of a
strong-coupling expansion of
$14^{\mbox{\begin{footnotesize}th\end{footnotesize}}}$ order. In
  VCA the ground state energy per lattice site is obtained from the
  grand potential $\Omega$ evaluated at the stationary point via the
  relation $E_0 = \Omega + n\,\mu$, and the one-particle density
  matrix $C(|\ve{r_i}-\ve{r_j}|)$ is evaluated from the Fourier
transform of the momentum distribution $n(\ve{k})$. As for the
spectral properties we obtain the best results for both the ground
state energy $E_0$ as well as the one-particle density matrix
$C(|\ve{r_i}-\ve{r_j}|)$ when using the variational parameters $
\mat{x}=\lbrace \mu,\,t,\,\delta \rbrace $ and obc for the reference
system. Therefore, from now on we restrict the calculations to this
set of variational parameters and obc for the reference system, and
evaluate the convergence properties of VCA with respect to the size of
the reference system. In \figc{fig:energy}{a}, we compare the VCA
results for the ground state energy $E_0$ for hopping parameters
corresponding to the first Mott lobe with the strong-coupling
results obtained from B.~Damski \textit{et~al.} in
Ref.~\onlinecite{damski_mott-insulator_2006}. 
We observe very good agreement between the two approaches. The deviations of the VCA results from the perturbative results are shown in \figc{fig:energy}{b}. For the largest cluster of $L=12$ sites the deviation is even less than $0.005$ for all values of the hopping strength $t$ corresponding to the first Mott lobe.

\begin{figure}
        \centering
        \includegraphics[width=0.48\textwidth]{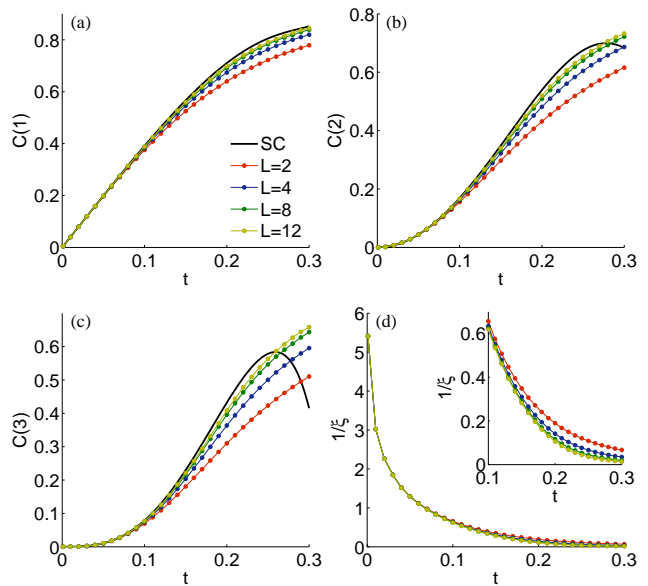}
        \caption{ (Color online) One-particle density matrix $C(|\ve{r_i}-\ve{r_j}|)$ for \fc{a} nearest neighbors $C(1)$, \fc{b} next nearest neighbors $C(2)$ and \fc{c} next-next nearest neighbors $C(3)$ obtained from VCA (dots connected by lines) and strong-coupling perturbation theory (solid line), respectively. \fc{d} shows the dominating exponential drop-off $1/\xi$ of the one-particle density matrix for large enough distances in dependence of the hopping strength $t$ and the size of the reference system. The inset shows an extract of $1/\xi$ where the hopping strength is restricted to $0.1\leq t \leq 0.3$.
	}
        \label{fig:corr}
\end{figure}

In Ref.~\onlinecite{damski_mott-insulator_2006} the one-particle density matrix for nearest neighbors $C(1)$, next nearest neighbors $C(2)$ and next-next nearest neighbors $C(3)$ have been evaluated as a function of the hopping strength $t$ by means of strong-coupling perturbation theory and the results have been compared with DMRG data evaluated for an $N=40$ site system. For $C(1)$ the strong-coupling results compare well with DMRG up to the hopping strength $t\approx 0.3$ and for $C(2)$ and $C(3)$ up to $t\approx 0.2$. In \fig{fig:corr} we compare our VCA results with the one-particle density matrix of Ref.~\onlinecite{damski_mott-insulator_2006} and find good agreement within the above mentioned regions of the hopping strength.

The one-particle density matrix decays exponentially for large enough distances $|\ve{r_i}-\ve{r_j}|$, \ie,
\begin{equation}
 C(|\ve{r_i}-\ve{r_j}|) \propto e^{-|\ve{r_i}-\ve{r_j}|/\xi}\;\mbox{,}
\end{equation}
where $\xi$ is the correlation length and its inverse describes the dominating exponential drop-off. The inverse of the correlation length $1/\xi$, shown in \figc{fig:corr}{d}, is almost zero when approaching the lobe tip which is already a precursor for the superfluid phase where the correlation length diverges.

\section{\label{sec:conclusion}Conclusions}
In the present work, we benchmarked the variational cluster approach by
means of the one-dimensional Bose-Hubbard model. In particular, we investigated the convergence properties of the variational cluster approach with respect to the variational parameter space, cluster size and boundary conditions of the reference system. For reference systems with obc we introduced, additionally to the chemical potential $\mu$ and the hopping strength $t$, the variational parameter $\delta$, which allows for a modified on-site energy at the boundary of the cluster. Using the additional on-site energy $\delta$ as variational parameter drastically improves the results as it 
restores to large extend the uniform particle density, which is violated by the breakup of the \psystem{} into decoupled
clusters. The resulting densities are essentially uniform, both that of  the reference system as well as that for the \psystem{} computed via the variational cluster approach Green's function without periodization.

At first we compared the variational cluster approach results for the phase boundary delimiting the first Mott lobe with essentially exact density matrix
renormalization group data.\cite{khner_one-dimensional_2000} Especially accurate results for the phase boundary have been obtained using the variational parameter set $\mat{x}=\lbrace \mu,\,t,\, \delta \rbrace$ and reference systems with obc, and $\mat{x}=\lbrace \mu,\,t \rbrace$ and reference systems with pbc. However, when extrapolating the results to infinitely large clusters the $\mat{x}=\lbrace \mu,\,t,\, \delta \rbrace$ variation with obc is superior to $\mat{x}=\lbrace \mu,\,t \rbrace$ and pbc.
Naively, one would expect to obtain better results if additional variational parameters are introduced and with  increasing
cluster size. Both is not generally true. For instance, 
augmenting the initial set of parameters $\mat{x} =\{\mu\}$ by
the intra-cluster hopping parameter $t'$ worsens the results. Similarly, using the parameter set 
$\mat{x} =\{\mu,t\}$ and increasing the cluster size leads to monotonically increasing deviations from the exact results.
This trend is accompanied by an increasing deviation of $t'$ from the value $t$ of the \psystem{}.
This poses the problem as to how to diagnose convergence toward the correct result, in cases where the latter is not known.
As an indication for correct results one may look at the deviations of the variational parameters at the stationary point of the grand potential from the \psystem{} parameters. These deviations are expected to shrink with increasing cluster size as larger clusters should better approximate the \psystem{}. For $\mat{x}=\lbrace \mu,\,t \rbrace$ and obc, however, the above considerations are not fullfilled and thus the poor results for the phase boundary can be ascribed to the failure of the criteria on the deviations of the variational parameters. In fact this criteria can be used to test the variational cluster approach results. Interestingly, for the $\mat{x}=\lbrace \mu,\,t \rbrace$ variation with obc we also observe increasing deviations of the cluster particle density from uniform distribution with increasing cluster size, which might indicate that boundary states play an important role for this configuration.
Single particle spectral functions evaluated with $\mat{x}=\lbrace \mu \rbrace$ and $\lbrace \mu,\,t \rbrace$ for reference systems with obc exhibit spurious gaps.\cite{koller_variational_2006} These spurious gaps are not visible anymore if the additional on-site energy $\delta$ is considered as variational parameter, which is a very important improvement. Spectral functions calculated for reference systems with pbc are not as smooth as those evaluated for reference systems with obc, but overall they exhibit the characteristic properties.

Second, we investigated the convergence properties of static
quantities, such as the ground state energy and the one-particle
  density matrix. We compared our variational cluster approach
results obtained for various values of the hopping strength $t$,
corresponding to the first Mott lobe, with results obtained by
means of high-order strong-coupling perturbation
theory.\cite{damski_mott-insulator_2006} We investigated the
convergence properties of the variational cluster approach using the
variational parameters $\mat{x}=\lbrace \mu,\,t,\, \delta \rbrace$ and
reference systems with obc, since this configuration yields the best
results for both the spectral properties as well as the static
properties. For the ground state energy we found excellent agreement
between the variational cluster approach results and the
strong-coupling results. Moreover the one-particle density
  matrix obtained by means of the variational cluster approach matches
very well with the strong-coupling results, which are for next nearest
neighbors and next-next nearest neighbors reliable for a hopping
strength $t \lesssim 0.2$. Finally, we evaluated the dominating
exponential decay of the one-particle density matrix, which is
the inverse correlation length. Close to the lobe tip the exponential
decay is almost zero, corresponding to an almost infinitely large
correlation length. This is already a precursor for the superfluid
phase. 

In summary, the variational cluster approach yields accurate results
with relatively low computational effort for both 
the phase boundary 
and the static properties of lattice bosons in one dimension, even at the
tip of the first Mott lobe where correlation effects are most
pronounced. 

\begin{acknowledgments}
We thank {H.~Monien} for providing us the DMRG data used in \figg{fig:spe:bhphasediagram}{fig:spe:bhphasediagrampbc}. M.K. is grateful to B.~Damski for fruitful discussions. We made use of parts of the ALPS library\cite{albuquerque_alps_2007} for the implementation of lattice geometries and for parameter parsing.
We acknowledge financial support from the Austrian Science Fund (FWF) under the doctoral program ``Numerical Simulations in Technical Sciences'' Grant No. W1208-N18 (M.K.) and under Project No. P18551-N16 (E.A.).
\end{acknowledgments}

\end{document}